\documentstyle[preprint,aps]{revtex}
\begin{document}
\draft
\preprint{RU9520}
\title{Spectroscopy of $^{194}$Po}

\author{W. Younes, J. A. Cizewski, H. -Q. Jin,~\cite{A}
L. A. Bernstein,~\cite{B} and D. P. McNabb}
\address{Department of Physics and Astronomy, Rutgers University,
New Brunswick, New Jersey 08903}

\author{C. N. Davids, R. V. F. Janssens, T. L. Khoo, C. J. Lister,
D. J. Blumenthal, M. P. Carpenter,
D. Henderson, R. G. Henry,~\cite{C} T. Lauritsen, D. T. Nisius,
H. T. Penttil\"a}

\address{Physics Division, Argonne National Laboratory,
Argonne, Illinois 60439}

\author{M. W. Drigert }

\address{
Idaho Nuclear Engineering Laboratory, Idaho Falls, Idaho 83415}

\date{PRC (Rapid Communications) \today}
\maketitle
\begin{abstract}
Prompt, in-beam $\gamma$ rays following the reaction
$^{170}$Yb + 142 MeV $^{28}$Si were measured at the
ATLAS facility using 10 Compton-suppressed Ge detectors
and the Fragment Mass Analyzer.  Transitions in
$^{194}$Po were identified and placed using $\gamma$-ray
singles and coincidence data gated on the mass of the
evaporation residues.  A level spectrum up to
J$\approx$10$\hbar$ was established.  The structure of $^{194}$Po
is more collective than that observed in the heavier polonium
isotopes and indicates that the structure has started to
evolve towards the more collective nature expected for
deformed nuclei.
\end{abstract}

\narrowtext

The onset of collective motion in nuclei near shell closures and
the mechanism by which this collectivity evolves towards deformation continue
to be open questions in nuclear physics.  While nuclei with valence
particles of only one type, such as Sn (Z=50) or Pb (Z=82)
isotopes, are not expected to become deformed in their ground
states, we now recognize that the structure of these nuclei as a
function of excitation energy and angular momentum can be associated with
a wide variety of shapes.  At moderate temperatures two-particle, two-hole
(2p-2h) proton excitations~\cite{1} across the shell gap give rise
to rotational structures in these nuclei.  At moderate angular momenta,
superdeformed shapes[2] have been identified in Pb isotopes.  In
nuclei with two valence proton holes, such as the Cd and Hg isotopes,
a similar coexistence between spherical vibrations, 4h-2p deformed
bands,~\cite{1} and superdeformed rotational bands (in
Hg)~\cite{2} are observed.  In contrast there is less
evidence~\cite{3} for the corresponding 4p-2h excitations in Te
isotopes, and relatively little is known about the collective motion
in the Po isotopes with two valence protons.~\cite{4}

Recently we reported on the first in-beam measurements in the
N=112 $^{196}$Po isotope~\cite{5}.  Delayed $\gamma$-rays
in $^{196}$Po had been initially studied by Alber, et al.~\cite{6},
who made spin and parity assignments based only on systematical
behavior.  In ref. ~\cite{5} we discussed $^{196}$Po as
a good example of a harmonic vibrator. The
evidence includes evenly-spaced energy levels up to the 8$^+$ state,
nearly degenerate members of the 2- and 3-phonon multiplets, and
branching ratios for the deexcitation of non-yrast levels which are
consistent with a simple vibrational picture.
However, the energy ratio R(4/2) = E(4$^+$)/E(2$^+$) is 1.92 in
$^{196}$Po, and is essentially constant for several isotopes.  Since
R(4/2) is expected to be 2.0 for a harmonic vibrator, the N$>$112 Po isotopes
are only weakly collective.  The
present work was proposed to investigate any possible change in shape in
the lighter Po isotopes.  By extending the systematics to a more
neutron-deficient isotope, we also hoped to confirm the
vibrational character of the non-yrast excitations in
$^{196,198}$Po.

The formation and subsequent study of the spectroscopy of such a
neutron-deficient, heavy nucleus is very difficult.  The statistical
evaporation code PACE~\cite{7} predicts that the most favorable
reaction, $^{28}$Si + $^{170}$Yb, at $\approx$ 142 MeV yields only a
$\approx$ 10 mb
cross section for $^{194}$Po, which competes with a 170 mb fission cross
section.  Furthermore, approximately 50\% of the $\gamma$-ray
intensity populating the low-lying states of $^{196}$Po in a
similar reaction~\cite{5} goes through an 850-ns 11$^-$ isomer.  If similar
conditions hold in $^{194}$Po, any prompt spectroscopy experiment
would miss about half of the already small cross-section, as well
as suffer from intense competition from fission.  Any successful
measurement of $^{194}$Po would, therefore, require some
additional selection criteria to enhance the channel of interest.

The present experiment to study $^{194}$Po was carried out at the
Argonne Tandem Linac Accelerator System (ATLAS) using the Fragment
Mass Analyzer~\cite{8} (FMA) to mass-select the evaporation residues.  A
beam of 142 MeV $^{28}$Si ions interacted with an enriched (70\%)
500~$\mu$g/cm$^2$ $^{170}$Yb target.  Prompt $\gamma$ rays were
recorded using an array of 10 Compton-suppressed Ge detectors located at
the target position.  The recoiling evaporation residues were
analyzed by the FMA, a triple-focusing mass separator which
disperses recoils by their mass/charge ratio.  A position-sensitive
parallel-plate avalanche counter (PPAC) located at the FMA focal plane
was used to determine the position and flight time of the recoils.
A large area silicon detector behind the PPAC
provided further energy information on the recoil products.  The
FMA was tuned so that both 14$^+$ and 15$^+$ charge states of A=194
recoils could be detected at the focal plane.  Single $\gamma$-ray
events were recorded to tape only when they occurred in
coincidence with a signal in the PPAC.  The energy and time information of
the Ge detectors were recorded to tape, together with the position
($\rm x_{PPAC}$), energy, and time-of-flight information measured at the
focal plane.  All double and higher fold $\gamma$-ray events were also
written to tape, without the requirement of a coincidence with a residue.
In total 71 x 10$^6$ $\gamma$-ray events were
recorded, of which only about 1\% occurred in coincidence with a
recoil at the PPAC.

The data were first sorted into an $\rm x_{PPAC}-\gamma$ matrix which was
used to identify $\gamma$ rays from A=194 nuclei.  Contamination
from the scattered primary beam products was reduced by setting an additional
two-dimensional gate on a recoil energy vs. time-of-flight matrix.
Figure~1 shows the projection of this $\rm x_{PPAC}-\gamma$ matrix onto
the position axis.  A portion of the $\gamma$-ray spectrum
corresponding to the A=194 mass gate with charge states 14$^+$ and 15$^+$
is shown in Fig.~2a.
When similar cuts were carried out on the other mass bins,
contamination from A$\neq$194 in the A=194 gate was found to be
negligible.  Discriminating between the different A=194 isobars
was a more difficult task.  Since the FMA resolution cannot
distinguish A=194 isobars and the predicted cross sections~\cite{7} for
$^{194}$Po and $^{194}$Bi are roughly comparable, as many as half
the recoils in the A=194 bin could correspond to Bi residues.  However, the
strongest lines assigned to $^{194}$Po are observed to be in coincidence
with Po x-rays.  We observe only one line, at 632 keV, strong enough
to be clearly seen in coincidence
with Bi x-rays and assign this transition to $^{194}$Bi.  That only one
strong line could be assigned to $^{194}$Bi
is not unexpected since it is an odd-odd nucleus and the
$\gamma$-ray strength is expected to be fragmented over many
decay paths.

The data were then sorted into a $\gamma$-$\gamma$ matrix gated on
the A=194 $\rm x_{PPAC}$ mass gate.  Because of the poor statistics, this
matrix was only used to confirm coincidences between transitions
clearly visible in the A=194 gated $\gamma$-ray spectrum.  Eight
transitions were identified and assigned to $^{194}$Po.  Five of
these, the 318.6(2), 365.7(2), 461.0(3), 544.6(3), and 600.7(4)
keV lines, were strong in the mass-gated $\gamma$-ray spectrum,
mutually coincident in the mass-gated $\gamma$-$\gamma$ matrix and
were, therefore, taken to be part of the ground state quasi-band.
They were ordered according to their intensity in the recoil-gated
$\gamma$-ray spectrum.  The sum of coincidence spectra gated on
these $\gamma$ rays is shown in Fig.~2b.  The remaining three
transitions - 340.3(3), 525.1(3), and 329.2(3) keV - were placed
using the $\gamma$-$\gamma$ data.  No angular information could be
extracted from the data due to poor statistics and, possibly, to
loss of alignment following de-excitation through short-lived
isomers.  Therefore, spin assignments were inferred from
systematics. The deduced level spectrum for $^{194}$Po in Fig.~3
summarizes our results.

The (2$^+_2$) and (4$^+_2$) assignments to the non-yrast levels at
659 and 1209 keV, respectively, are based on the systematical behavior
of the levels in the heavier isotopes.  These states are too low in
excitation to be negative-parity levels, but their energies are consistent
with the positive-parity non-yrast levels seen in $^{196,198}$Po.
A 659-keV transition cannot be assigned from our data displayed
in Fig. 2a.  This is consistent with the branching ratio for
the deexcitation of the (2$^+_2$) state in $^{196}$Po,
as reported in ref. ~\cite{5}.  Similarly, from the decay pattern in
$^{196}$Po~\cite{5}, it would be difficult to observe a transition
between the (2$^+_2$) and (4$^+_2$) states in $^{194}$Po.

The level structure of $^{194}$Po shows a clear difference from
that of the heavier, weakly collective isotopes.  First, the
R(4/2) ratio is 2.15 in $^{194}$Po, and the spacings of the other
yrast levels are no longer the equal spacings characteristic of a
harmonic vibrator.  The systematics of the energy ratios are
displayed in Fig.~4a, which clearly shows the R(6/4) ratio greater
than expected for a harmonic vibrator.  However, the essentially vibrational
character of the lighter Po isotopes appears to be preserved in $^{194}$Po.
As in the heavier isotopes, the 2$^+_2$ and 4$^+_2$ states lie very close
in energy to the 4$^+_1$ and 6$^+_1$ levels, respectively, as is expected
with nearly degenerate members of the two- and three-phonon multiplets.
If the non-yrast 2$^+_2$ and 4$^+_2$ states were members of a 4p-2h
configuration, their excitation energies should have dropped as
the middle of the neutron shell is approached and their energy
separation should have decreased.  Instead, this spacing has
slightly increased, and the relative energy differences of these
non-yrast states follow closely those of the corresponding member
of the vibrational multiplet.

What is unexpected is that the evolution of the collectivity in
the polonium isotopes is different from that of other isotopes
with two valence protons.  To illustrate this feature we
have displayed in Fig.~4b the change in energy of the 2$^+_1$
states, normalized to the average 2$^+$ energy.  The data for the Po isotopes
with N~$\leq$~126 are compared to those for the Cd and Te isotopes
with N~$\leq$~82 and the Hg isotopes with N~$\leq$~126 and identical
numbers of valence neutrons, N$_{\rm n}$.  For these Cd, Te, and Hg isotopes
the change in the 2$^+$ energies as a function of N$_{\rm n}$ is within
10\% over a wide range of isotopes away from the closed neutron shell.
While the change for many of the Po
isotopes is also relatively small, for the lightest isotopes,
$^{194,196}$Po, the change is very large:  about 140 keV, with
respect to the heavier isotope.  If the 2$^+$ energy in $^{192}$Po
did drop another 140 keV with respect to $^{194}$Po, the 2$^+$ energy
would be close to the value at which the phase transition from
anharmonic vibrator to deformed rotor occurs for a larger number
of nuclei with 38 $<$ Z $<$ 82~\cite{9}. This would represent a rapid change
to a more collective, possibly deformed structure of the N $<$ 110
isotopes, a change not seen in other regions of the periodic
table.

In conclusion we have identified for the first time excited states
in $^{194}$Po.  This nucleus has the character of an anharmonic
vibrator, with non-yrast states following the expectations of
members of phonon multiplets.  However, the evolution of the shape
seems to be much more rapid than is observed in other two-proton
isotopes.  The systematical behavior of the Po isotopes suggests
that $^{192}$Po could be a critical nucleus:  it could either
be deformed if the drop in 2$^+$ energies continues or the shape
could stabilize at the moderate collectivity of $^{194}$Po.

This work was supported in part by the National Science Foundation
and Department of Energy under contracts W-31-109-ENG-38 and
DE-AC07-76ID01570.

\begin{figure}
\caption{Projection of the recoil-$\gamma$ matrix onto the
position axis, $\rm x_{PPAC}$.  The mass assignments of the peaks are given.}
\end{figure}

\begin{figure}
\caption{(a) Projection of the $\rm x_{PPAC}-\gamma$ matrix onto the
$\gamma$-ray energy axis.  The insert displays the x-ray region of
the spectrum from the $\gamma$-$\gamma$ matrix gated on the
A=194 mass and 632-keV line.(b)  Sum of spectra gated on transitions from all
yrast states (2$^+$, 4$^+$, 6$^+$, 8$^+$, 10$^+$) from a $\gamma$-$\gamma$
matrix gated on the A=194 mass.  The insert displays the x-ray region of
the spectrum from the $\gamma$-$\gamma$ matrix gated on the
A=194 mass and 319-keV line.}
\end{figure}

\begin{figure}
\caption{Deduced level scheme of $^{194}$Po.  Spin assignments
are based on systematics.}
\end{figure}

\begin{figure}
\caption[]{(a)  R(4/2) and R(6/4) ratios for levels in Po (closed
symbols) and Te (open symbols). The ratios for the harmonic vibrator
are indicated. (b)  The difference in 2$^+$
energies [$\Delta$E= E(A) - E(A-2)] divided by the average 2$^+$ energy
[$<$E$>$ = \{E(A) + E(A-2)\}/2] as a function of the number of valence
neutrons, N$_{\rm n}$, for Po with 110~$\leq$~N~$\leq$~126 (closed
symbols) and Cd, Te with 66 $\leq$ N $\leq$ 82 and Hg with
104~$\leq$~N~$\leq$~126 (open symbols) nuclei.
Data are taken from Refs.~\cite{4,5,6} and the present work.}
\end{figure}

\begin{references}
\bibitem[*]{A} Present address: Oak Ridge National Laboratory, Oak Ridge, TN
37831 USA.
\bibitem[\dag]{B} Present address: Lawrence Livermore National Laboratory,
Livermore, CA 94550 USA.
\bibitem[\ddag]{C} Present address: University of California - San Francisco
Medical Center, San Francisco, CA  USA.
\bibitem{1} J.L. Wood, K. Heyde, W. Nazarewicz, M. Huyse, and P. van
Duppen, Physics Reports {\bf 215}, 101 (1992), and references
therein.
\bibitem{2} B. Singh and R. B. Firestone, ``Table of Superdeformed
Rotational Bands,'' LBL-35916 (1994).
\bibitem{3} C.S. Lee, {\it et al}., Nucl. Phys. {\bf A528}, 381 (1991), and
Nucl. Phys. {\bf A530}, 58 (1991); and J.A. Cizewski, L. A. Bernstein,
R.G. Henry, H.-Q. Jin, C.S. Lee, and W. Younes, in {\it Capture
Gamma-ray Spectroscopy}, J. Kern (ed)., World Scientific (1994), p.
328.
\bibitem{4} Evaluated Nuclear Structure Data Files, Brookhaven National
Laboratory, Upton, NY.
\bibitem{5} L.A. Bernstein {\it et al}., Phys. Rev. {\bf C52}, (1995, in
press).
\bibitem{6} D. Alber {\it et al}., Z. Phys. {\bf A339}, 225 (1991).
\bibitem{7} Computer code PACE written by A.
Gavron, Phys. Rev. {\bf C21}, 230 (1980) and modified by J. Beene.
\bibitem{8} C.N. Davids {\it et al}., Nucl. Instrum. Methods {\bf
B70}, 358 (1992).
\bibitem{9} R.F. Casten, N.V. Zamfir, and D.S. Brenner, Phys. Rev.
Lett. {\bf 71}, 227 (1993).
\end{references}
\end{document}